\documentclass[runningheads]{llncs}
\usepackage[T1]{fontenc}
\usepackage{graphicx}
\begin{document}
\title{Explainability for Embedding AI: Aspirations and Actuality}
%
%

\author{
    Thomas Weber
    \email{thomas.weber@ifi.lmu.de}
    \orcidID{0000-0002-6894-605X}
}
%
\authorrunning{Weber}
%
\institute{LMU Munich, Munich, Germany}
\maketitle              
\begin{abstract}
With artificial intelligence (AI) embedded in many everyday software systems, effectively and reliably developing and maintaining AI systems becomes an essential skill for software developers. However, the complexity inherent to AI poses new challenges. Explainable AI (XAI) may allow developers to understand better the systems they build, which, in turn, can help with tasks like debugging.
In this paper, we report insights from a series of surveys with software developers that highlight that there is indeed an increased need for explanatory tools to support developers in creating AI systems. However, the feedback also indicates that existing XAI systems still fall short of this aspiration. Thus, we see an unmet need to provide developers with adequate support mechanisms to cope with this complexity so they can embed AI into high-quality software in the future.
\keywords{explainable AI \and explanatory debugging \and debugging AI \and data-driven development.}
\end{abstract}
\section{Introduction}
In recent years, the proliferation of artificial intelligence (AI) technologies has revolutionized various industries, with practical applications in fields such as medicine \cite{Shehab2022}, marketing \cite{Brei2020}, IT security \cite{Apruzzese2023}, and also in everyday life.
However, building reliable systems in many of these domains is already challenging.
Adding the complexity inherent in AI often makes the software hard to understand for developers and end-users alike. This is particularly the case because, with these types of applications, some critical behavior is no longer encoded by the developer but is instead inferred from data. This makes debugging and maintenance of these systems increasingly challenging.

Explainable AI (XAI) has emerged as a promising domain to address some of these challenges by adding transparency and interpretability to the AI models and their behavior. Ideally, this would mean that developers can use these XAI mechanisms to understand the software they write and any opaque behavior that may be inferred from the data. This might help with debugging and thus enhance the overall robustness and reliability of AI-powered software systems.

However, despite its potential benefits, XAI is not without its limitations. Aside from technical challenges, like negatively affecting  performance~\cite{Berk2017,Biswas2020}, the interpretability of AI models may vary depending on factors such as model complexity, data heterogeneity, and task characteristics, posing challenges to the generalizability of XAI solutions. Furthermore, how much value an XAI system can provide can be subjective, raising concerns about the benefit across different users and applications.
Thus, creating XAI systems not with a technical perspective but relying on human-centered research to determine the actual practical benefits, is critical for their success.

In this paper, we outline some insights we have collected through surveys with software developers regarding XAI systems and how much they actually benefit from them. They highlight that, indeed, XAI system could be a way to support developers but also that current popular XAI systems may still fall short of this aspiration. This emphasizes that there is still a continuing need for human-centric research on XAI methods beyond their initial case studies.


\section{Related Work}

While embedding AI into software opens up many opportunities for novel applications, it also introduces new challenges. Since in modern AI systems, the developer no longer encodes the behavior of the system, but instead, it is inferred from data, it becomes challenging to understand their behavior, particularly when things do not go as intended.

In response, XAI research tries to make these systems more explainable, interpretable, and understandable~\cite{GUNNING2019}. Over the years, people have developed many mechanisms, visualizations, and tools to provide explanations in some form on another
\cite{Adadi2018XAI,barbalau2020generic,Mengnan2019XAI,goebel2018xai,HOFFMANN2018ARXIV,lundberg2019}. Some popular examples for these are LIME~\cite{LIME} which uses an explanatory surrogate model or SHAP~\cite{SHAP}, which utilizes Shapley values to assign importance to input features.

The idea to use explanatory systems for debugging software has also been around for quite some time, initially for traditional software~\cite{5635185}, but increasingly also specifically targeting AI systems~\cite{10.1145/2678025.2701399,9533944} due to their complexity and opacity.

The research community has, however, acknowledged several shortcomings of current XAI systems~\cite{10.1145/3334480.3383047}. For example, while there are many viable use cases, many current XAI systems insufficiently address users' needs in these scenarios. One reason for this is their often prototypical nature, but also the fact that these XAI systems are very narrow in scope and focus only on use cases from their conception~\cite{10.1145/3334480.3383047}. Thus, more human-centric research will be necessary.

\section{Surveys}

To determine how software developers perceive XAI systems, we collected feedback through three online surveys.

\paragraph{Survey 1: The Need for Explainability}

In the first survey, our goal was to initially gauge the need for explainability in general. At this point, we considered a broad opinion from developers and end users alike. To this end, we applied the survey scale described by Weber et al.~\cite{weber2021interact} with a series of different scenarios and types of software. In the initial version of the survey, we included nine applications that embed AI as one part of its functionality. Five were picked to be widely available (online search, social media, multimedia platforms, shopping recommender systems, and navigation), and four were not yet as common (autonomous driving, predictive policing, robotics, personalized medicine). These examples were based on common examples for AI applications from the literature~\cite{Cohen2019,EIBAND2019,GADE2019,goebel2018xai,holzinger2017xai,ROBOTICS,WANG2019}.

The survey started with a consent form and questions about the participants' demographic background and technology affinity.
Then, each scenario and application was briefly described to participants, after which they answered statements about their understanding of and interest in explanations for this system using five-point Likert scales.
To put the demand for explainability of AI systems into context, we later added an additional five applications that do not utilize AI for their core functionality (file management, web browser, email, office software like MS PowerPoint, media manipulation software like Adobe Photoshop). The order of the applications was randomized in the survey.

\paragraph{Survey 2: Explainability for Developers}
Based on the results of the first survey, we decided to investigate further how XAI methods are perceived by software developers, specifically with the goal of improving the development experience and understanding the systems they implement.

After initial consent and demographic items, we collected feedback on existing XAI systems.
We used LIME~\cite{LIME} and SHAP~\cite{SHAP} as exemplary systems, as these two are also popular and established examples in the literature~\cite{10.1145/3334480.3383047}.
For each of these, participants received a brief explanation and a concrete example using several datasets~\cite{HARRISON197881}.
After familiarizing themselves with the XAI system, participants responded to a series of statements about it using five-point Likert scales. These statements were based on the survey scales on the demand for explainability from the previous survey. Still, they were extended and rephrased to accommodate the presence of a dedicated explanatory system and focus on the perspective of software developers.
After this, we tested the participants' understanding of the presented methods by asking them a series of multiple-choice questions. Each of these was phrased as a potential conclusion that could have been drawn from the XAI system. Participants then needed to decide whether the conclusion was drawn from the XAI system.
As before, the order in the survey was randomized.

\paragraph{Survey 3: Explainability for Recognizing Faults}

Finally, the third survey aimed to see whether XAI allowed developers to draw their own conclusions to detect issues in the system.

To determine whether participants could utilize XAI methods to detect faults, we asked them to use again LIME and SHAP the California Housing Dataset~\cite{KELLEYPACE1997291}. In this survey, we manipulated the dataset to introduce faults. To simulate skewed data, we replaced the median income in a quarter of the data points where the median house value was above 200.000. Additionally, we swapped the labels of one of the most important features, the longitude, and one of the least important, the number of households. Since latitude remained an important feature, this led to semantic inconsistencies, as the position of a household consisted of one very important and one seemingly unimportant feature.

Again, we first collected participants' consent and demographic information. After this, the dataset and the explanatory methods were introduced. We asked participants to investigate whether the system works as intended using LIME and SHAP in separate study conditions. Each participant used the unmodified and the modified dataset. The order of the conditions was randomized.
They then answered questions about their experience with the XAI system and the dataset in six categories: (1) ranking of the three most important features, (2) multiple choice questions about the relationship between features and prediction, as well as five-point Likert scales on (3) their subjective understanding of the model, (4) the perceived plausibility of the explanation of the model, and (5) their confidence in the model. Finally, we questioned (6) whether they could detect anomalies or errors in the data or the model.

\subsection{Apparatus and Participants}
We conducted all surveys online via an institutional survey platform. Participants could take the survey at any time but were required to complete it in one continuous session.
The link to the survey and, thus, recruitment was distributed through mailing lists of our institution and external contacts with professional developers. Participants were compensated with the equivalent of 10 \$US per hour.

This recruitment yielded 96 participants for the first survey with just AI-powered systems and 116 who completed the full survey with AI-powered and traditional examples (94 male, 112 female, six other). Half of this sample were students of various fields with an average age of 25 years (SD: 5.7 years).
The second survey was completed by 17 participants (12 male, four female, and one other), either Computer Science students or working as data scientists (3), software engineers or architects (5), or in research (1).
Finally, the third survey was completed by 21 participants (12 male, 6 female, 3 other). For this survey, we had a majority of responses from Computer Science students (14), while the other seven participants were professional software developers (3), data scientists (2), or academic researchers (2). In consequence half of the participants was younger than 25 while the other half was in the 25--34 year range.

\section{Results}

In the following, we will summarize highlights from the survey responses.

In the first survey, we observed a generally high demand for explanations for all 14 presented systems.
However, taking into account technology affinity, we observed that participants reported low technology affinity had an equally high demand for explainability for AI and non-AI applications. Participants with higher technology affinity and particularly background in Computer Science had on average a lower demand for explanations for the traditional applications. However, for the AI-powered systems, their demand for explanations was, on average, about the same as that of the inexperienced participants.

As this suggests, even experienced users may require explanations for software with AI functionality about as much as the average user. The goal of our analysis was then to see whether XAI systems can assist them in this situation. However, based on the feedback from the second survey, this does not appear to be the case immediately.
In the responses, we observed quite consistently equal groups of those who considered the XAI systems positively, helpful, and supporting their understanding, as well as those who were rather negative and saw no immediate benefit from them. The only area where there was a notable imbalance was regarding trust in the system, where two-thirds of the participants saw no positive impact of the XAI systems for calibrating their trust.

When taking into account experience, this picture shifts, though: for example, participants with multiple years of data science experience were generally more confident in their ability to interpret explanatory output, thus also seeing the greater benefit of understanding the behavior of the AI system. Similarly, the more experienced participants found the XAI systems generally easy to understand while only a single less experienced participant considered them easy to understand.

Also, less experienced participants were all uncertain to sceptical whether LIME or SHAP helps them detect issues with an AI system they are building. Meanwhile, half of the experienced developers saw at least some benefit for finding faults, while the other half was neutral for both XAI systems.

Thus, we used the third survey to test the potential benefit of finding faults.
Feedback from this survey indicates that participants had an overall positive impression of the XAI methods for their given task. The perceived understanding, subjective plausibility, and confidence in the model were overall high, with no significant differences between the two XAI methods. However, only two participants were able to detect one of the deliberately added faults when using SHAP, and only one participant found a fault using LIME.

\section{Discussion}

Considering the feedback from these surveys, it is quite clear that developers are a viable target group for XAI. For end-users, explanations may be useful, but that is true, regardless of whether the system has embedded AI functionality. From an end-user's perspective, the internal mechanisms are opaque either way. A mental model where the presence of AI makes a difference will often require at least some degree of technical expertise.

Software developers bring expertise and typically willingness to engage with the details of AI. Understanding existing traditional systems is already well supported because software developers can benefit from their computational thinking skills~\cite{wing} to construct an adequate mental model. Additionally, there are well-established debugging tools, logging output, etc.
However, these do not seem to satisfy the needs of experienced users. XAI may be a solution here to support developers.

However, the popular XAI tools we presented in our survey, but also many more, appear to be designed more towards a specific initial use case and less for generally exploring potential faults in an AI system or the underlying data. The skewed responses of the experienced users might also suggest that XAI merely helps support existing suspicions or highlight known potential issues. Experienced users may already know what they might be looking for and use the XAI system to confirm their hypotheses. While this can be useful, care needs to be taken that developers do not over-rely on XAI just to support their existing beliefs. The fact that XAI systems seem to be built only for the experts themselves has also been criticized before~\cite{inmates} and means that novice users have an even harder time figuring out faults in their systems.

Even so, the actual rate at which participants could detect deliberately added faults was not great either. One can imagine that even more obscure or situational errors are even harder to detect. However, it must be noted that the XAI systems in our study were not designed as dedicated debugging tools but to get a general understanding of the AI models. At the same time, though, the example use cases we presented were not particularly complex either, and still, the participants had trouble finding the added faults. The participant sample with many inexperienced users may also be a reason for this. However, this only further emphasizes the previous point that these tools require prior knowledge to be useful in the first place.

\section{Conclusion}

To ensure that software with embedded AI functionality is reliable and generally of high quality it is essential that software developers have appropriate tools to understand what is going on. XAI has the potential to be one such tool that assists developers in understanding and debugging their applications. However, popular XAI systems are not necessarily engineered with such a goal in mind. As feedback from our surveys suggests, they currently require a good deal of prior knowledge to be beneficial, and even then, it is unclear whether they simply support existing mental models or actually provide a broader benefit for finding unknown issues in AI systems.

The feedback underlines that even for experienced developers AI systems pose a challenge and how limited the tool support still is. At the same time, developers are a target group that could benefit particularly much from XAI systems. This makes XAI for developers a particularly interesting research area that can help developers and, by extension, improve the quality of AI applications in general.

\bibliographystyle{splncs04}
\bibliography{main}

\end{document}